%Paper: hep-ph/9404240
%From: wise@theory.caltech.edu (Mark Wise)
%Date: Fri, 8 Apr 94 12:03:24 PDT

\input phyzzx
% Titlepage macros
\hoffset=0.2truein
\voffset=0.1truein
\hsize=6truein
\def\TITLEPAGE{\frontpagetrue}
\def\CALT#1{\hbox to\hsize{\tenpoint \baselineskip=12pt
        \hfil\vtop{
        \hbox{\strut CALT-68-#1}}}}

\def\ABSTRACT#1{\vskip .2in \vfil \centerline{\twelvepoint
\bf Abstract}
        #1 \vfil}
\def\ENDTITLEPAGE{\vfil\eject\pageno=1}

\tolerance=10000
\hfuzz=5pt

%\doublespace %= for phys. rev. papers
\TITLEPAGE
\CALT{1928}         %if title takes 2 lines use \break
\bigskip
\titlestyle {Remark on Charm Quark Fragmentation to $D^{**}$ Mesons\foot{Work
supported in part by the U.S. Dept. of Energy
under Grant no. DE-FG03-92-ER 40701.}}
\bigskip
\smallskip
\centerline{Yu-Qi Chen}
\centerline{{\it China Center for Advanced Science and Technology}}
\centerline{{\it (World Laboratory) P.O. Box 8730, Beijing 100080, China}}
\bigskip
\centerline{Mark B. Wise}
\centerline{{\it California Institute of Technology, Pasadena, CA  91125}}
\bigskip
\bigskip
\ABSTRACT{The observed $D^{**}$ mesons have $c\bar q$ flavor quantum numbers
and spin-parity of the light degrees of freedom $s_\ell^{\pi_{\ell}} = 3/2^+$.
In the $m_c \rightarrow \infty$ limit the spin of the charm quark is conserved
and the $c \rightarrow D^{**}$ fragmentation process is characterized by the
probability for the charm quark to fragment to a $D^{**}$ meson with a given
helicity for the light degrees of freedom.  We consider the calculated $b
\rightarrow B_c^{**}$ fragmentation functions in the limit $m_c/m_b \rightarrow
0$ as a qualitative model for the $c \rightarrow D^{**}$ fragmentation
functions.  We find that in  this model charm quark fragmentation to
$s_\ell^{\pi_{\ell}} = 3/2^+$ light degrees of freedom with helicities $\pm
1/2$ is favored over fragmentation to $s_\ell^{\pi_{\ell}} = 3/2^+$ light
degrees of freedom with helicities $\pm 3/2$.}
\ENDTITLEPAGE

\eject

Heavy quark spin-flavor symmetry$^{[1,2]}$ has important consequences for the
spectroscopic properties$^{[3]}$ of hadrons containing a single heavy quark
$Q$.  In the $m_Q \rightarrow \infty$ limit the spin of the heavy quark $\vec
S_Q$ is conserved.  It is convenient for classifying states to introduce the
spin of the light degrees of freedom $\vec S_\ell = \vec S - \vec S_Q$.  In the
$m_Q \rightarrow \infty$ limit hadrons containing a single heavy quark $Q$ are
labeled by the quantum number $s_\ell$ and these states come in degenerate
doublets with total spins $s_\pm$ that arise from combining the spin of the
light  degrees of freedom with that of the heavy quark
$$	s_\pm = s_\ell \pm 1/2 . \eqno (1)$$
(An exception is the case $s_\ell = 0$ where there is only a single hadronic
state with total spin $1/2$.)

Recently Falk and Peskin noted that heavy quark spin symmetry relates
fragmentation probabilities for members of these hadron doublets.$^{[4]}$  It
implies that the probability, $P_{h_{Q} \rightarrow s, h_{s}}^{(H)}$, for a
heavy quark $Q$ with spin along the fragmentation axis $h_Q$ to fragment to a
hadron $H$ in a doublet with spin of the light degrees of freedom $s_\ell$,
total spin $s$ and total spin along the fragmentation axis (i.e., helicity)
$h_s$ is
$$	P^{(H)}_{h_{Q} \rightarrow  s, h_{s}} = P_{Q \rightarrow s_{\ell}}
p_{h_{\ell}} |< s_Q, h_Q; s_\ell, h_\ell | s, h_s>|^2 ,\eqno (2)$$
where $h_\ell = h_s - h_Q$.  In eq. (2) $P_{Q \rightarrow s_{\ell}}$ is the
probability for the heavy quark to  fragment to a doublet with spin of the
light degrees of freedom $s_\ell$.  It is independent of the heavy quark spin
but will depend on other quantum numbers needed to specify the doublet
containing $H$.  $p_{h_{\ell}}$ is the conditional probability that the light
degrees of freedom have helicity $h_\ell$ (given that $Q$ fragments to
$s_\ell$).   The probability interpretation implies that $0 \leq p_{h_{\ell}}
\leq 1$ and

$$	\sum_{h_{\ell}} p_{h_{\ell}} = 1 . \eqno (3)$$
Parity invariance of the strong interactions implies that
$$	p_{h_{\ell}} = p_{-h_{\ell}} . \eqno (4)$$
Eqs. (3) and (4) restrict the number of independent probabilities
$p_{h_{\ell}}$ to be equal to $s_\ell - 1/2$ for mesons and $s_\ell$ for
baryons.  At the hadron level parity invariance of the strong interactions
implies that
$$	P_{h_{Q} \rightarrow s, h_{s}}^{(H)} = P_{-h_{Q} \rightarrow s,
-h_{s}}^{(H)}  . \eqno (5)$$
Heavy quark spin symmetry reduces the number of independent fragmentation
probabilities.  For mesons with spin of the light degrees of freedom $s_\ell$
the fragmentation probabilities $P_{h_{Q} \rightarrow s, h_{s}}^{(H)}$ are
expressed in terms of the $s_\ell - 1/2$ ($s_\ell$ for baryons) independent
$p_{h_{\ell}}$'s and $P_{Q \rightarrow s_\ell}$.

For the $D$ and $D^*$ mesons $s_\ell^{\pi_{\ell}} = 1/2^-$ and eqs. (3) and (4)
imply that $p_{1/2} = p_{-1/2} = 1/2$.  The relative fragmentation
probabilities.
$$P_{1/2 \rightarrow 0,0}^{(D)} \quad : \quad P_{1/2 \rightarrow 1,1}^{(D^{*})}
\quad : \quad  P_{1/2 \rightarrow 1,0}^{(D^{*})} \quad : \quad P_{1/2
\rightarrow 1, -1}^{(D^{*})}  , \eqno (6)$$
are
$$ 1/4 \qquad : \qquad 1/2 \qquad : \qquad 1/4 \qquad : \qquad 0 \qquad . \eqno
(7)$$
Eq. (5) determines the fragmentation probabilities for a helicity $-1/2$ charm
quark in terms of those above.  The relative fragmentation probabilities for
the three $D^*$ helicities agree with experiment, however, the prediction that
the probability for a charm quark to fragment to a $D$ is $1/3$ the probability
to fragment to a $D^*$ does not.  Part of the origin of this violation of heavy
quark symmetry may be in the $D^* - D$ mass difference which suppresses
fragmentation to the $D^*$.

Excited spin-one and spin-two charmed mesons $D_1$(2420) and $D_2^*$(2460) have
been observed experimentally.  Their properties suggest that they are members
of a doublet with spin-parity of the light degrees of freedom
$s_\ell^{\pi_{\ell}} = 3/2^+$.  These states are sometimes referred to as
$D^{**}$ mesons.  For this multiplet eqs. (3) and (4) imply that there is only
one independent conditional probability $p_{h_{\ell}}$.  Falk and Peskin take
it to be $w_{3/2}$, the conditional probability to fragment to helicities $\pm
3/2$, and write
$$	p_{3/2} = p_{-3/2} = {1\over 2} w_{3/2} , \eqno (8a)$$
$$	p_{1/2} = p_{-1/2} = {1\over 2} (1 - w_{3/2}) . \eqno (8b)$$
Eq. 2 implies that the relative fragmentation probabilities$^{[4]}$
$$	P_{1/2 \rightarrow 1,1}^{(D_{1})} ~~ : ~~ P_{1/2 \rightarrow 1,0}^{(D_{1})}
{}~~ : ~~ P_{1/2 \rightarrow 1,-1}^{(D_{1})} ~~ : $$
$$	P_{1/2 \rightarrow 2,2}^{(D_{2}^{*})} ~~ : ~~ P_{1/2 \rightarrow
2,1}^{(D_{2}^{*})} ~~ : ~~ P_{1/2 \rightarrow 2,0}^{(D_{2}^{*})} ~~ : ~~ P_{1/2
\rightarrow 2, -1}^{(D_{2}^{*})} ~~ : ~~ P_{1/2 \rightarrow 2,
-1}^{(D_{2}^{*})} , \eqno (9)$$
are
$$	{1\over 8} (1 - w_{3/2})~~:~~{1\over 4} (1 - w_{3/2}) ~~:~~{3\over 8}
w_{3/2} ~~:$$
$$	{1\over 2} w_{3/2} ~~:~~ {3\over 8} (1 - w_{3/2})~~:
{}~{1\over 4} (1 - w_{3/2}) ~~:~~ {1\over 8} w_{3/2} ~~:~~ 0 .\eqno (10)$$
Total fragmentation probabilities are given by multiplying eq. (10) by $P_{c
\rightarrow 3/2}$.

Eq. (10) predicts that the ratio of $D_1$ to $D_2^*$ production by charm quark
fragmentation is $3/5$, independent of $w_{3/2}$.  Assuming that $D^{**}$
decays are dominated by $D^{(*)} \pi$ final states the experimental value of
this ratio is close to unity$^{[5]}$ (with an error at the 20\% level).
However, Eichten, et al., have suggested that $D^{**} \rightarrow D^{(*)}
\pi\pi$ decays are also important.$^{[6]}$  Experimentally the probability to
fragment to helicity $\pm 3/2$ light degrees of freedom is small, i.e.,
$w_{3/2} \lsim 0.24$ at the 90\% confidence level.$^{[4]}$  In this brief
report we show that constituent $q\bar q$ pair production by a virtual gluon is
a fragmentation process that produces a small value for $w_{3/2}$.

The fragmentation functions for $b \rightarrow B_c^{**}$ were computed in Ref.
[7] using perturbative QCD.  They are of order $\alpha_s^2 (m_c)$ and arise
from a single virtual gluon producing a $c\bar c$ pair.  Taking the limit
$m_c/m_b \rightarrow 0$, of the results in Ref. [7] for fragmentation to the
spin-2 $B_c^{**}$ state, gives $^{[8]}$
$$	w_{3/2} = 29/114 . \eqno (11)$$
Eq. (11) is our result.  The probability to fragment to helicities $\pm 3/2$ is
roughly one-third the probability to fragment to helicities $\pm 1/2$.  This
suppression may arise because fragmentation to helicity $\pm 3/2$ light degrees
of freedom requires a non-zero opening angle with respect to the fragmentation
axis to conserve total angular momentum.

The value of $w_{3/2}$ in eq. (11) arose from perturbative QCD.  Of course the
$c \rightarrow D^{**}$ fragmentation process is nonperturbative.  Nonetheless,
we find it interesting that a simple physical mechanism can give a small value
for $w_{3/2}$.  If the value of $w_{3/2}$ is measured to be roughly equal to
0.25 then production of a constituent $q\bar q$ pair by a virtual gluon may
provide a reasonable qualitative picture for the $c \rightarrow D_2^{**}$
fragmentation process.
\medskip
\centerline{{\bf References}}

\item{1.}  N. Isgur and M.B. Wise, Phys. Lett., {\bf B232} 113 (1989); Phys.
Lett., {\bf B237}, 527 (1990).

\item{2.}  See also:  S. Nussinov and W. Wetzel, Phys. Rev., {\bf D36} 130
(1987); M.A. Shifman and M.B. Voloshin, Sov. J. Nucl. Phys., {\bf 47} 511
(1988).

\item{3.}  N. Isgur and M.B. Wise, Phys. Rev. Lett., {\bf 66} 1130 (1991).

\item{4.}  A.F. Falk and M.E. Peskin, SLAC-PUB-6311 (1993) unpublished.

\item{5.}  P. Avery, et al., (CLEO collaboration), CLNS 94/1280 (1994)
unpublished.

\item{6.}  E.J. Eichten, et al., FERMILAB-Pub-93/255-T (1993) unpublished.

\item{7.}  Yu-Qi Chen, Phys. Rev., {\bf D48} 5181 (1993).

\item{8.}  There are several typographical errors in eqs. (60) and (61) of Ref.
[7].  The correct expression for $W_2$ in eq. (60) is minus that given in Ref.
[7] and in eq. (61) the correct expression for $W_0^0$ is $W_0^0 = (z^2
\alpha_2^2 - 2 z^2 \alpha_2 - 2 z \alpha_2 + z^2 + z + 1)^2 (1 - z)/12$.

\bye